\documentclass[3p,times,twocolumn]{elsarticle}

\usepackage{ecrc}

\volume{00}

\firstpage{1}

\journalname{Nuclear Physics B Proceedings Supplement}

\runauth{}

\jid{nuphbp}

\jnltitlelogo{Nuclear Physics B Proceedings Supplement}

\usepackage{amssymb}
\usepackage[figuresright]{rotating}

\begin{document}

\begin{frontmatter}



\dochead{}

\title{Status of Leading-Order Hadronic Vacuum Polarization Dispersion Calculation}


\author{Zhiqing Zhang}

\address{Laboratoire de l'Acc\'el\'erateur Lin\'eaire, Univ.\ Paris-Sud 11 et IN2P3/CNRS, France}

\begin{abstract}
The leading-order hadronic contribution to the muon magnetic anomaly $a_\mu\equiv (g_\mu-2)/2$, calculated using a dispersion integral of $e^+e^-$ annihilation data and $\tau$ decay data, is briefly reviewed. This contribution has the largest uncertainty to the predicted value of $a_\mu$, which differs from the experimental value by $\sim 3.6$ (2.4) standard deviations for the $e^+e^-$ ($\tau$) based analysis. New results since the last workshop and main open issues on the subject are discussed.
\end{abstract}

\begin{keyword}
muon magnetic anomaly \sep hadronic vacuum polarization \sep $e^+e^-$ annihilation \sep tau spectral function

\end{keyword}

\end{frontmatter}

% \linenumbers

\section{Introduction}
\label{sec:intro}

The Standard Model (SM) has been extremely successful. The only missing particle of the SM, the Higgs boson, may have been discovered recently at the LHC, once verified with more data. All SM predictions have been tested often to an extraordinary precision and no sign of new physics has been found with few exceptions. One such exception is the well known muon $g-2$ anomaly, $a_\mu$. The status as of the Tau 2010 workshop is about 3.6 standard deviations between the direct measurement dominated by the E821 experiment at BNL~\cite{bnl06} and the corresponding SM predictions~\cite{Hoecker:2010qn}.

The SM prediction $a^{\rm SM}_\mu$ is usually decomposed into three parts
\begin{equation}
a_\mu^{\rm SM}=a^{\rm QED}_\mu+a^{\rm weak}_\mu+a^{\rm had}_\mu\,,
\end{equation}
corresponding to QED, weak and hadronic loop contributions, respectively. The dominant QED contribution includes all photonic and leptonic $(e, \mu, \tau)$ loops starting with the classic $\alpha/2\pi$ Schwinger contribution. It has been computed recently through 5 loops and has the following numerical value~\cite{qed}:
\begin{equation}
a_\mu^{\rm QED}=(11\,658\,471.8951\pm 0.0080)\times 10^{-10}\,. 
\end{equation}

The weak part includes loop contributions involving heavy $W^\pm$, $Z$ and Higgs particles. It is suppressed by at least a factor $\alpha/\pi\cdot m^2_\mu/M^2_W\simeq 4\times 10^{-9}$. The numerical value accounting for the dominant 1- and 2-loop contributions~\cite{weak1,weak2,weak3,weak4} is
\begin{equation}
a_\mu^{\rm weak}=(15.4\pm0.1\pm0.2)\times 10^{-10}\,,
\end{equation}
where the uncertainties stem from quark triangle loops and the assumed Higgs mass range between 100 and 500\,GeV, which may be reduced based on the preliminary Higgs mass determination at the LHC.

The hadronic part involving quark and gluon loop contributions may be further decomposed into leading-order (LO), higher-order (HO) and light-by-light (LBL) scattering contributions $a_\mu^{\rm had}=a_\mu^{\rm had,\, LO}+a_\mu^{\rm had,\, HO}+a_\mu^{\rm had,\, LBL}$. At present, the LO contribution cannot reliably be calculated from perturbative QCD (pQCD) and is determined instead by a dispersion relation~\cite{dispersion}
\begin{equation}
a_\mu^{\rm had,\, LO}=\frac{1}{3}\left(\frac{\alpha}{\pi}\right)^2\int^\infty_{m^2_{\pi^0\gamma}}ds\frac{K(s)}{s}R^{(0)}(s)\,,
\end{equation}
where $R^{(0)}(s)$ represents  the ratio of the bare cross sections of $e^+e^-$ annihilation into hadrons to the point-like muon-pair cross section and  $K(s)\sim 1/s$ is a QED kernel function~\cite{kqed} and gives a strong weight to low-energy part of the integrand. The precision of $a_\mu^{\rm had,\, LO}$ depends thus on that of the $e^+e^-$ annihilation data in particular that of $\rho(770)\to \pi^+\pi^-$ and it has the largest uncertainty to $a_\mu^{\rm SM}$ and this is why most of the effort from both experimental and theoretical sides went into its improved determination over the last 20 years or so.

In the following, we shall briefly describe the new development since the last workshop and discuss a few open issues on the subject.

\section{New development and open issues}
The preliminary DHMZ\,10 results shown at the Tau 2010 workshop have been published (this and all following numbers are given in units of $10^{-10}$)~\cite{dhmz10}:
\begin{equation}
a_\mu^{\rm had,\, LO}=692.3\pm 1.4\pm 3.1\pm 2.4\pm 0.2\pm 0.3\label{eq:amuee}
\end{equation}
where the first error is statistical, the second channel-specific systematic, the third common systematic, correlated between at least two exclusive channels, and fourth and fifth errors stand for the narrow resonance and QCD uncertainties, respectively. For this new $e^+e^-$ based prediction, we included new $\pi^+\pi^-$ cross section data from KLOE, all available multi-hadron data from BABAR, a reestimation of missing low-energy contributions using results on cross sections and process dynamics from BABAR, a reevaluation of all experimental contributions using the software package HVPTools together with a reanalysis of inter-experiment and inter-channel correlations, and a reevaluation of the continuum contributions from pQCD at four loops. The new result is 3.2 below the previous one~\cite{dhmyz09}. This shift is composed of $-0.7$ from the inclusion of the new, large photon angle data from KLOE, $+0.4$ from the use of preliminary BABAR data in the $e^+e^-\to \pi^+\pi^-2\pi^0$ mode, $-2.4$ from the new high-multiplicity exclusive channels, the reestimate of unknown channels, and the new resonance treatment, $-0.5$ from mainly the 4-loop term in the QCD prediction of the hadronic cross section as well as smaller other differences. There was a minor update for the FF 2012 workshop~\cite{ffworkshop} by including the latest BABAR $2\pi^+2\pi^-$, $2K2\pi$ and $2K2\pi^0$ channels resulting in $a_\mu^{\rm had,\, LO}=692.4\pm 1.3\pm 3.1\pm 2.3\pm 0.2\pm 0.3$.

The $\pi^+\pi^-$ channel used to be limited in precision, so it was proposed in~\cite{adh98} to transform the corresponding tau spectral function through an isospin rotation to the $e^+e^-$ cross section by $\sigma^{l=1}\left(e^+e^-\to \pi^-\pi^-\right)=4\pi\alpha^2/s\cdot v(\tau^-\to \pi^-\pi^0\nu_\tau)$ and to provide an independent evaluation after accounting for all isospin breaking effects~\cite{detal09}.
Similar transformations can be made for four-pion channels. The resulting tau based prediction reads
\begin{equation}
a_\mu^{\rm had,\ LO}[\tau]=701.5\pm 3.5\pm 1.9\pm 2.4\pm 0.2\pm 0.3
\end{equation}
where the first error is $\tau$ experimental, the second the uncertainty of isospin-breaking corrections~\cite{detal09}, the third $e^+e^-$ experimental, and the last two the narrow resonance and QCD uncertainties. The $2\pi$ and $4\pi$ channels account for about 78\% of the LO hadronic contribution, the rest is taken from the $e^+e^-$ channels or pQCD calculations.

Adding to these results the contributions from $a_\mu^{\rm had,\, HO}=-9.84\pm 0.07$~\cite{hlmnt11}, computed using a similar dispersion relation approach, $a_\mu^{\rm had,\, LBL}=10.5\pm 2.6$~\cite{lbl}, estimated from theoretical model calculations, as well as $a_\mu^{\rm QED}$ and $a_\mu^{\rm weak}$, one gets 
\begin{eqnarray}
a_\mu^{\rm SM}[e^+e^-]\!\!&=&\!\!11\,659\,180.2\pm 4.9_{\rm tot}\,,\label{eq:ee}\\
a_\mu^{\rm SM}[\tau]\!\!&=&\!\!11\,659\,189.4\pm 5.4_{\rm tot}\,.\label{eq:tau}
% \hspace{2mm} (4.9_{\rm tot})
\end{eqnarray}
%where the uncertainties correspond to LO and HO hadronic, and other contributions, respectively. 
The $e^+e^-$ ($\tau$) based prediction deviates from the direct experimental average~\cite{bnl06} of
\begin{equation}
a_\mu^{\rm exp}=11\,659\,208.9\pm 5.4_{\rm stat}\pm 3.3_{\rm syst}
\end{equation}
by $28.7\pm 8.0$ $(19.5\pm 8.3)$, i.e.\ $3.6\,\sigma$ $(2.4\,\sigma)$.

\begin{figure}[htb]
\begin{center}
\vspace{-2mm}
\includegraphics[width=.475\textwidth]{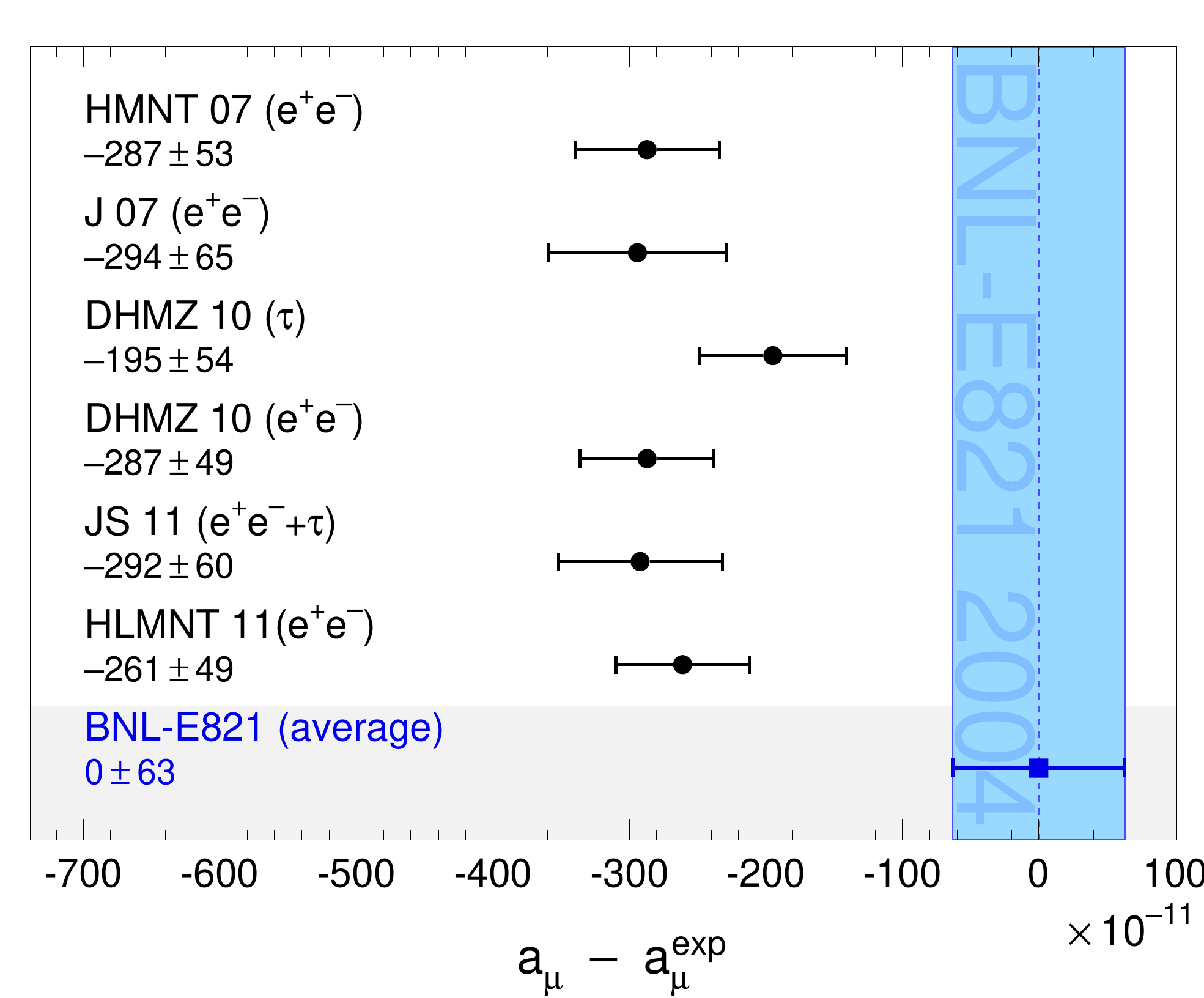}
\end{center}\vspace{-7mm}\caption{Compilation of recent results for $a_\mu^{\rm SM}$ (in units of $10^{-11}$), subtracted by the central value of the experimental average. The shaded vertical band indicates the experimental error.}
\label{fig:amures}
\end{figure}

A compilation of recent $a_\mu^{\rm SM}$ predictions in comparison with the experimental average of direct measurements is shown in Fig.\,\ref{fig:amures}. In particular the prediction of HLMNT\,11~\cite{hlmnt11} is similar to that of DHMZ\,10 (Eq.(\ref{eq:amuee})). The input $e^+e^-$ data sets used are largely identical. They differ mainly in the data combination and error treatment. This is reflected in Table~\ref{tab:comp} (extracted from Table~4 in~\cite{hlmnt11}). The difference is comparable to or larger than one of the quoted errors. In addition, the quoted errors are quite different. It is desirable that these differences can be understood and reduced in the future. 
\begin{table}[htb]
\begin{center}
\begin{tabular}{crrr}
Channel & HLMNT\,11 & DHMZ\,10 & diff. \\\hline
$K^+K^-$ & $22.09\pm 0.46$ & $21.63\pm 0.73$ & 0.46 \\
$\pi^+\pi^-$ & $505.65\pm 3.09$ & $507.80\pm 2.84$ & $-2.15$ \\
$\pi^+\pi^-\pi^0$ & $47.38\pm 0.99$ & $46.00\pm1.48$ & 1.38\\\hline
\end{tabular}
\caption{Comparison for hadronic contributions to $a_\mu$ in the energy range from 0.305 to 1.8\,GeV from three $K^+K^-$, $\pi^+\pi^-$ and $\pi^+\pi^-\pi^0$ channels, extracted from Table~4 in~\cite{hlmnt11}.}
\label{tab:comp}
\end{center}
\end{table}   

\begin{figure}[htb]
\begin{center}
\includegraphics[width=.475\textwidth]{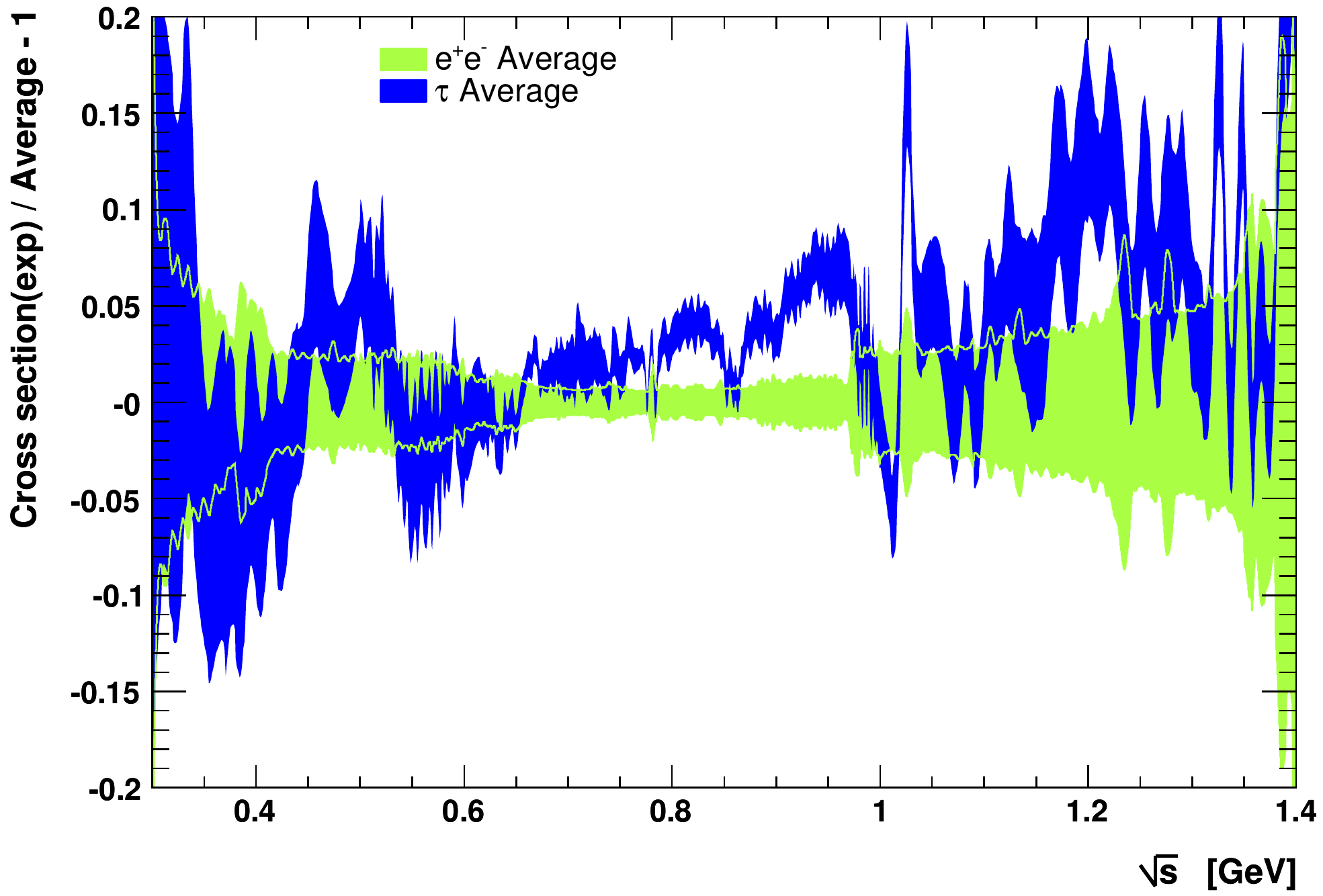}
\end{center}\vspace{-7mm}\caption{Relative shape comparison between ALEPH-Belle-CLEO-OPAL combined $\tau$ (dark shaded) and $e^+e^-$ spectral function (light shaded).}
\label{fig:ee_tau}
\end{figure}

The difference of $9.1\pm 5.0$, i.e.\ $1.8\,\sigma$ between the $\tau$ and $e^+e^-$ based predictions shown in Eqs.(\ref{eq:ee}) and (\ref{eq:tau}),  is one of the open issues. Jegerlehner and Szafron claim that the difference can be explained by the $\rho^0-\gamma$ mixing missing in the $\tau$ data~\cite{js11}. It remains to be checked whether this is the real explanation or there are experimental issues related to the $e^+e^-$ and $\tau$ measurements. Indeed, the $e^+e^-$ and $\tau$ difference can be seen from the relative shape comparison in the energy range between 0.3 and 1.4\,GeV in Fig.\,\ref{fig:ee_tau}.

\begin{figure}[htb]
\begin{center}
\includegraphics[width=.475\textwidth]{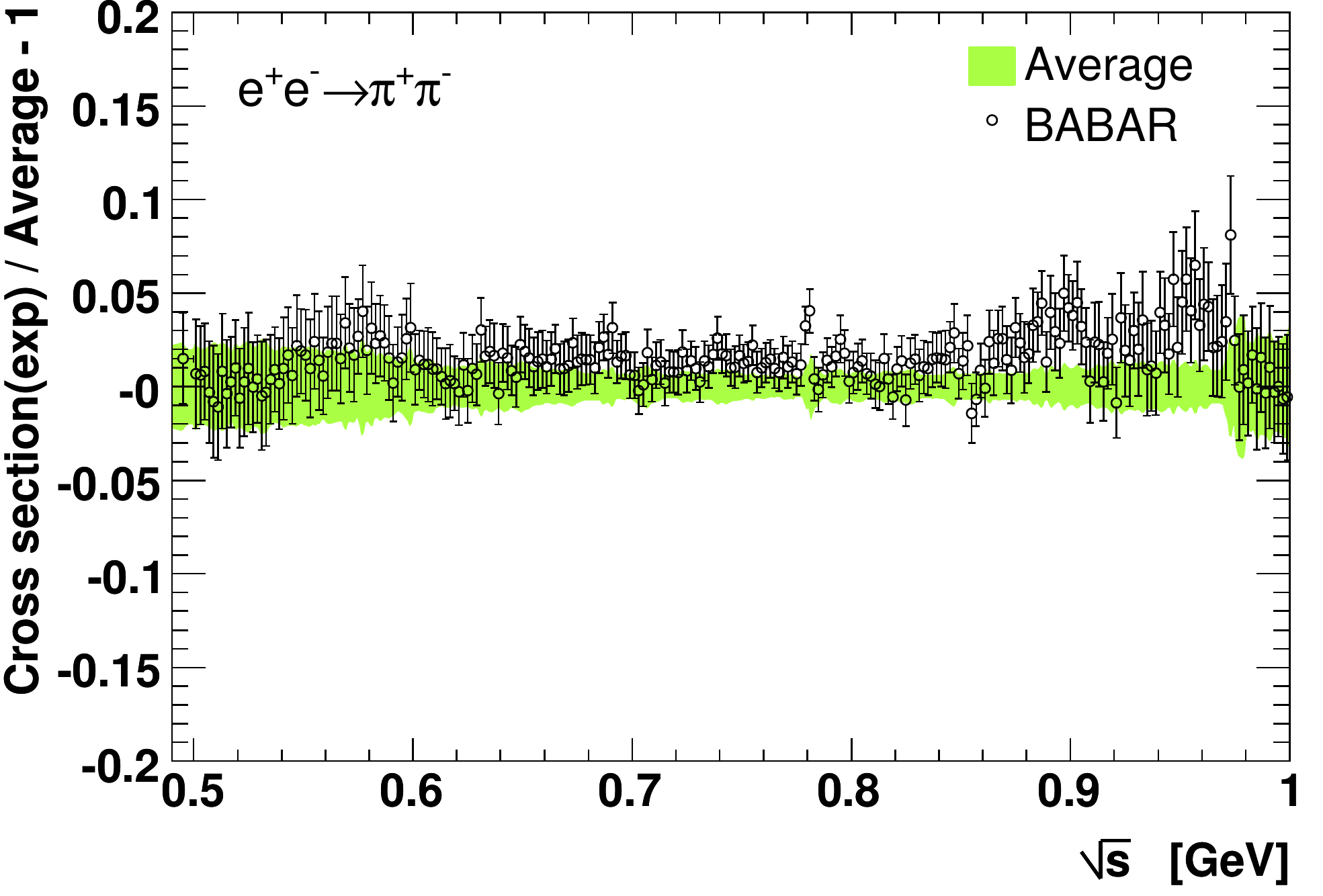}
\includegraphics[width=.475\textwidth]{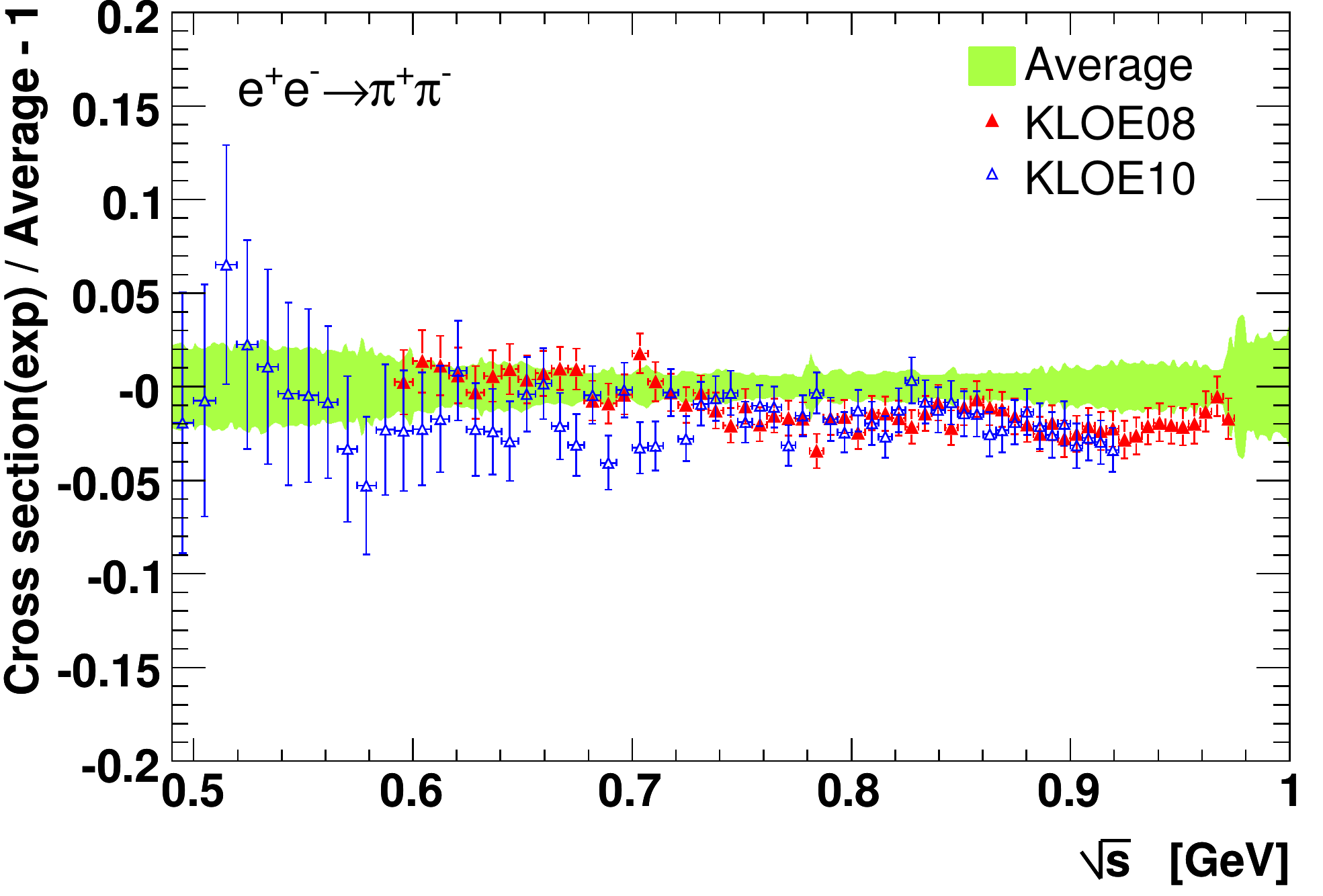}
\end{center}\vspace{-7mm}\caption{Comparison between individual $e^+e^-\to \pi^+\pi^-$ cross section measurements from BABAR (top) and KLOE (bottom) and the HVPTools average.}
\label{fig:babar_kloe}
\end{figure}

The other related issue is the different shape between BABAR and KLOE $\pi^+\pi^-$ cross section data (Fig.\,\ref{fig:babar_kloe}). This difference, leading to an amplified uncertainty in the combination following the PDG prescription, prevents further error reduction. The published KLOE measurements were still performed without involving the ratio of pion-to-muon pairs as BABAR did. It is known that some of the systematic uncertainties cancel in the latter ratio measurement.

\begin{figure}[htb]
\begin{center}
\includegraphics[width=.475\textwidth]{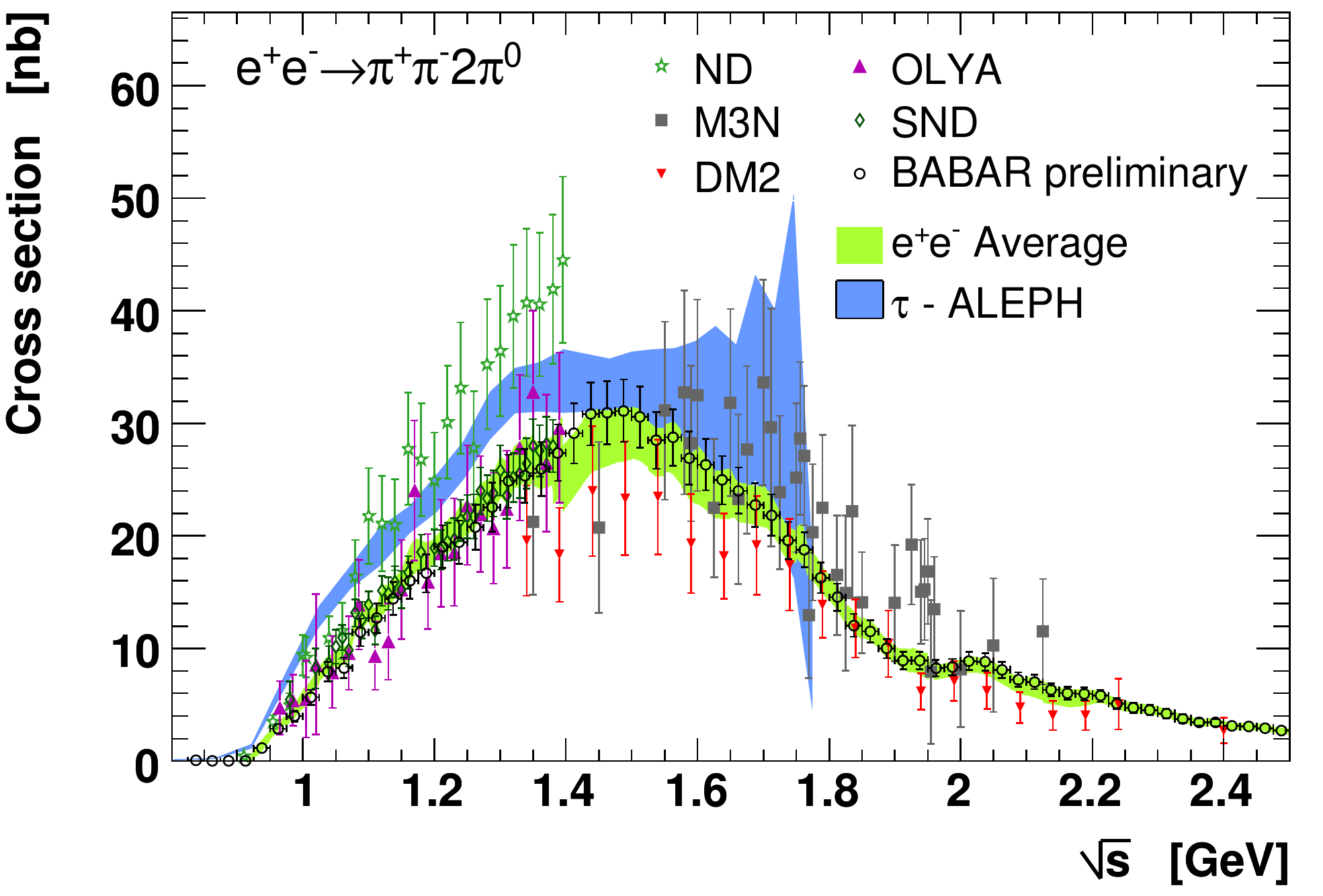}
\end{center}\vspace{-7mm}\caption{Cross section of $e^+e^-\to 2\pi^+2\pi^0$ versus center-of-mass energy. The shaded green (blue) band gives the HVPTools average for $e^+e^-$ ($\tau$) data.}
\label{fig:2pi2pi0_ee_tau}
\end{figure}

Another problematic channel concerns $e^+e^-\to \pi^+\pi^-2\pi^0$ (Fig.\,\ref{fig:2pi2pi0_ee_tau}). There is a large scattering between measurements from different experiments, in particular between ND and other experiments. In addition when comparing the $e^+e^-$ average with the $\tau$ average, there is a significant difference in normalization. This discrepancy deserves further studies and clarification.

\section{Running $\alpha(s)$ at $M_Z^2$}
The running electromagnetic fine structure constant, $\alpha(s)=\alpha(0)/(1-\Delta_{\rm leo}(s)-\Delta_{\rm had}(s))$, at $s=M^2_Z$, is an important ingredient of the SM fit to electroweak precision data at the $Z$ pole. Similar to $a_\mu$, the error on $\alpha(M^2_Z)$ is dominated by hadronic vacuum polarization.

The sum of all the hadronic contributions gives for the $e^+e^-$ based prediction~\cite{dhmz10}:
\begin{equation}
\Delta \alpha_{\rm had}(M^2_Z)=(275.0\pm 1.0)\times 10^{-4}\,,\label{eq:alpha}
\end{equation}
 which is, contrary to the evaluation of $a_\mu^{\rm had,\,LO}$, not dominated by the uncertainty in the low energy data, but by contributions from all energy regions, where both experimental and theoretical errors are of similar magnitude. This is to be compared with a recent update by HLMNT~\cite{hlmnt11}: 
%and by adding the top quark contributions of $\Delta\alpha_{\rm top}=(-0.728\pm 0.014)\times 10^{-4}$~\cite{alphatop} to give: 
$\Delta\alpha_{\rm had}(M^2_Z)=(275.5\pm 1.4)\times 10^{-4}$.
 
The reduced electromagnetic coupling strength at $M_Z$ obtained in Eq.(\ref{eq:alpha}) leads to an increase by 7\,GeV in the central value of the Higgs boson mass obtained by the standard Gfitter fit~\cite{gfitter} to electroweak precision data, compared to the previous determination.

\section{Summary and perspectives}
The deviation of about 3.6\,$\sigma$ between the direct measurement and the SM predictions on $a_\mu$ is significant but not sufficient for claiming new physics. The $a_\mu$ deviation and the large $H\to\gamma\gamma$ rate observed at the LHC can however be explained by a light stau contribution~\cite{gps12}.

We have mentioned a few open issues in the current $e^+e^-$ data and the comparison between the $e^+e^-$ and $\tau$ data, in particular in the $\pi^+\pi^-$ and $\pi^+\pi^-2\pi^0$ channels. The $\pi^+\pi^-$ discrepancy between BABAR and KLOE in some of the energy ranges prevents us from achieving a better precision in the data combination. In order to significantly improve the uncertainty of the leading-order hadronic contribution, these issues need to be resolved by either more precise new measurements or better theoretical understandings. Lattice calculations are making significant progress, but are not yet competitive with the dispersion approach with data~\cite{blum}. The uncertainty of the light-by-light scattering contribution is the next item to improve.

The uncertainty of the direct measurement (dominated by the statistical precision) is now larger than the total uncertainty of the SM predictions. Two new $g-2$ experiments from Fermilab and JPARC are being built and an error reduction by a factor of 4 is expected from these experiments in a few years from now. It will be a challenge for new SM predictions to match this new level of accuracy.

\vspace{5mm}
{\small I am grateful to the fruitful collaboration with my colleagues and friends Michel Davier, Andreas Hoecker and Bogdan Malaescu.}
 


%% References

\nocite{*}
\bibliographystyle{elsarticle-num}
\bibliography{zhangz_tau12}






% \end{thebibliography}

\end{document}